\documentclass[%
 reprint,
pra,
]{revtex4-2}
\usepackage{amsmath,amssymb}
\usepackage{mathrsfs} 
\usepackage{graphicx}
\usepackage{physics}
\usepackage{setspace}
\usepackage[english]{babel}
\usepackage{cancel}
\usepackage{array}
\usepackage{mathtools}
\usepackage{my_symbols}

\begin{document}
\title{First-order coherent quantum Zeno dynamics and its appearance in tight-binding chains}
\author{Yuhan Mei}
\UR
\date{\today}

\begin{abstract}
     The coherent quantum Zeno dynamics (QZD) is a special unitary time evolution in which a quantum population transition gets constrained in a subspace of the entire Hilbert space. We show that coherent QZD can be categorized by orders for the first time, where only the zeroth-order type has been well investigated. In this paper, we focus on the little-known first-order coherent QZD (FC-QZD). We also construct some chain-like systems described by the tight-binding model which establishes FC-QZD in the form of a surprisingly nonlocal end-to-end population transition.
\end{abstract}
\maketitle
\section{Introduction}
Zeno's paradox is often introduced with phrases such as ``a watched flying arrow is motionless" or ``a watched pot never boils" \cite{Aristotle}. The quantum Zeno dynamics (QZD) was defined by Facchi \etal in the early 2000s \cite{facchiQuantumZenoDynamics2000,facchiQuantumZenoSubspaces2002}, which is a generalization of the more well-known quantum Zeno effect (QZE) \cite{misraZenoParadoxQuantum1977,peresZenoParadoxQuantum1980}. Both terms describe the consequence of ``watching" a quantum system. However, QZE refers to that, ``watching" stops the time evolution of its quantum state; whereas QZD refers to that, ``watching" constrains the evolution in a subspace in Hilbert space that includes the initial state. The ``watching" is sometimes interpreted as frequent measurements which induce decoherence \cite{facchiQuantumZenoSubspaces2002,facchiUnificationDynamicalDecoupling2004}. In this paper, we focus on another kind, called coherent QZD by us, where ``watching" is interpreted as a strong coupling described by a Hamiltonian \(H_w\), which rather keeps the evolution coherent \cite{facchiQuantumZenoSubspaces2002,facchiUnificationDynamicalDecoupling2004}. Coherent QZD has been experimentally realized on different platforms, such as rubidium BEC \cite{schaferExperimentalRealizationQuantum2014}, Rydberg atoms \cite{signolesConfinedQuantumZeno2014}, photons in a microwave cavity \cite{bretheauQuantumDynamicsElectromagnetic2015}, and trapped ions \cite{linPreparationEntangledStates2016}.

Now, let us explicate the elements that are necessary to establish coherent QZD. Suppose that quantum evolution used to be governed by a Hamiltonian \(H\). After adding the ``watching", the total Hamiltonian reads 
\al{
H_\text{tot}=\lambda^{-1}H_w+H=\lambda^{-1}(H_w+\lambda H)
\label{eq:lambdaH}
}
in Hilbert space \(\mathcal{H}_\text{tot}\) with \([H_w,H]\neq0\) \cite{facchiQuantumZenoSubspaces2002,facchiQuantumZenoSubspaces2003}. The factor \(\lambda\in[0,1]\) is employed to adjust the ``power of watching", i.e., the strength of the strong coupling. When \(\lambda\) is sufficiently small, the term in parentheses is subject to perturbation theory, which we will apply soon. Meanwhile, we suppose \(H_w\ket{\psi(0)}=0\) where \(\ket{\psi(0)}\) is the initial state. This assumption corresponds to the characteristic counter-intuitive feature of QZE and QZD \cite{homeConceptualAnalysisQuantum1997}: although ``watching" neither \textit{directly} facilitates nor inhibits the transition away from the initial state, it significantly alters the time evolution in the long term. For example, when Misra and Sudarshan coined the term ``quantum Zeno effect", they considered stopping an unstable particle's decay by continuously tracking its decay products, which never directly affects the initially undecayed particle state \cite{misraZenoParadoxQuantum1977}. In the same vein, \(H_w\) should have no direct effect on \(\ket{\psi(0)}\), i.e., \(H_w\ket{\psi(0)}=0\).

In the next section, we will employ perturbation theory to show why \(H_\text{tot}\) can lead to QZD characterized by constrained dynamics. A similar proof was once given by Facchi \etal without considering \(H_w\ket{\psi(0)}=0\) \cite{facchiQuantumZenoSubspaces2003}. We will then see that QZD can be categorized by orders which has not been done before. In fact, previous research has been considering the zeroth-order type, while we are going to investigate the little-known first-order coherent QZD (FC-QZD), which surprisingly manifests itself as nonlocal end-to-end population transition in some chain-like systems described by the 1-D tight-binding model, called tight-binding chains by us.

\section{From degenerate perturbation theory to coherent QZD}
\label{sec:perturb}

When \(\lambda\) is sufficiently small, \((H_w+\lambda H)\) in \Eq{eq:lambdaH} is subject to perturbation theory. The unperturbed and the perturbed Hamiltonians have eigenfunctions
\al{
H_w\ket{\phi_{n\alpha}^{(0)}}&=\eta^{(0)}_{n}\ket{\phi_{n\alpha}^{(0)}},\\
(H_w+\lambda H)\ket{\phi_{n\alpha}}&=\eta_{n\alpha}\ket{\phi_{n\alpha}}.
}
where \(n=0,1,2...\) labels energy levels, and the Greek alphabets label degeneracy. Since we have defined \(H_w\ket{\psi(0)}=0\), there exists \(\eta^{(0)}_{0}=0\). Also, we suppose that \(\eta^{(0)}_{0}\) has degeneracy, whose necessity will show up soon. For future use, relevant perturbative expansions are defined as
\al{
&{\eta}_{n\alpha}=\eta^{(0)}_{n}+\lambda\eta^{(1)}_{n\alpha}+\lambda^2\eta^{(2)}_{n\alpha}+\mathcal{O}(\lambda^3), \label{eq:eta}\\
&\ket{\phi_{n\alpha}}=\ket{\phi_{n\alpha}^{(0)}}+\lambda\ket{\phi_{n\alpha}^{(1)}}+\mathcal{O}(\lambda^2)\\
&P_{n\alpha}\equiv\ket{\phi_{n\alpha}}\bra{\phi_{n\alpha}}=P^{(0)}_{n\alpha}+\lambda P^{(1)}_{n\alpha}+\mathcal{O}(\lambda^2),\label{eq:proj}\\
&=\ket{\phi_{n\alpha}^{(0)}}\bra{\phi_{n\alpha}^{(0)}}+\lambda\smlb{\ket{\phi_{n\alpha}^{(1)}}\bra{\phi_{n\alpha}^{(0)}}+\ket{\phi_{n\alpha}^{(0)}}\bra{\phi_{n\alpha}^{(1)}}}+\mathcal{O}(\lambda^2),\nonumber
}
where the superscripts label the orders and \(P_{n\alpha}\) is the eigenprojector.

Since we are interested in the dynamics given by \Eq{eq:lambdaH}, we decompose the corresponding evolution operator as 
\begin{align}
\begin{aligned}
e^{-iH_\text{tot}t}=e^{-i\lambda^{-1}t(H_w+\lambda H)}&=e^{-i\tau\sum_{n\alpha}\smlb{{\eta}_{n\alpha}P_{n\alpha}}}\\&=\sum_{n\alpha}\smlb{e^{-i{\eta}_{n\alpha}\tau}{P}_{n\alpha}},
\end{aligned}
\label{eq:useries}
\end{align}
where \(\tau\equiv t/\lambda\). The derivation of last equality is shown in \footnote{The derivation of the last equality in \Eq{eq:useries}:\begin{align*}
&e^{-i\tau\sum_{n\alpha}\smlb{{\eta}_{n\alpha}P_{n\alpha}}}\\=&\prod_{n\alpha}\midb{1-i\tau{\eta}_{n\alpha}{P}_{n\alpha}+\dfrac{\smlb{-i\tau{\eta}_{n\alpha}{P}_{n\alpha}}^2}{2}+...}\\
=&\prod_{n\alpha}\midb{1+{P}_{n\alpha}\smlb{e^{-i\eta_{n\alpha}\tau}-1}}\\
=&1+\sum_{n\alpha}{P}_{n\alpha}\smlb{e^{-i\eta_{n\alpha}\tau}-1}\\
=&\sum_{n\alpha}\smlb{e^{-i{\eta}_{n\alpha}\tau}{P}_{n\alpha}},
\end{align*}
where we have used the normalization \(\sum_{n\alpha}P_{n\alpha}=1\) and the orthogonality \(P_{n\alpha}P_{m\beta}=\delta_{m,n}\delta_{\alpha,\beta}P_{n\alpha}\)
}. Then, substituting the projectors in the summation with \Eq{eq:proj} yields
\al{
\begin{aligned}
    e^{-iH_\text{tot}t}=\sum_{n\alpha}\smlb{e^{-i{\eta}_{n\alpha}\tau}P^{(0)}_{n\alpha}}+\mathcal{O}(\lambda).
\end{aligned}
\label{eq:uprime}
}
Next, we times \(\ket{\psi(0)}\) to both sides in order to investigate the time evolution of the state. Fortunately, the assumption \(H_w\ket{\psi(0)}=0\) enables some simplification in this step, because it implies \(\ket{\psi(0)}\in P^{(0)}_{0}\mathcal{H}_\text{tot}\), where \(P^{(0)}_{0}\equiv\sum_\alpha P^{(0)}_{0\alpha}\) is the projector into the space composed of all degenerate eigenstates of \(H_w\) regarding to \(\eta^{(0)}_0=0\). This means \(P^{(0)}_{n\neq0\alpha}\ket{\psi(0)}=0\), which simplifies the first term on the right-hand side of \Eq{eq:uprime} acting on \(\ket{\psi(0)}\):
\als{
\sum_{n\alpha}\smlb{e^{-i{\eta}_{n\alpha}\tau}P^{(0)}_{n\alpha}}\ket{\psi(0)}&=\sum_{\alpha}\smlb{e^{-i{\eta}_{0\alpha}\tau}P^{(0)}_{0\alpha}}\ket{\psi(0)}\nonumber\\
&=e^{-it\lambda^{-1}\sum_{\alpha}\smlb{{\eta_{0\alpha}P^{(0)}_{0\alpha}}}}\ket{\psi(0)},
}
where the second equality applies the reverse of \footnotemark[\value{footnote}]. By defining \(H_{\text{QZD}}\equiv \lambda^{-1}\sum_{\alpha}\smlb{{\eta_{0\alpha}P^{(0)}_{0\alpha}}}\), the entire \Eq{eq:useries} yields
\al{
e^{-iH_\text{tot}t}\ket{\psi(0)}=e^{-iH_\text{QZD}t}\ket{\psi(0)}+\mathcal{O}(\lambda)\ket{\psi(0)}.
\label{eq:timeop}
}
It is easy to verify \([H_{\text{QZD}}, P_0^{(0)}]=0\). Combining \(\ket{\psi(0)}\in P^{(0)}_{0}\mathcal{H}_\text{tot}\), we infer that the evolution \(e^{-iH_\text{QZD}t}\ket{\psi(0)}\) is restricted within the subspace \(P_0^{(0)}\mathcal{H}_\text{tot}\). The resulting constrained dynamics is what we call coherent QZD. To emphasize, only if \(\text{dim}(P^{(0)}_0)\geq 2\), \(P^{(0)}_{0}\mathcal{H}_\text{tot}\) can have enough space to place a transition from the initial state to somewhere else. This confirms the necessity of degeneracy in \(\eta^{(0)}_0=0\). Nonetheless, \(\mathcal{O}(\lambda)\ket{\psi(0)}\) transitions the population away from the subspace \(P_0^{(0)}\mathcal{H}_\text{tot}\) (proved in \App{sec:validity}). Therefore, \Eq{eq:timeop} suggests that the evolution due to \(H_\text{tot}\) is approximately coherent QZD, whereas some population at an order of \(\lambda^2\) leaks out of the subspace \(P_0^{(0)}\mathcal{H}_\text{tot}\).

\section{Orders in coherent QZD}
\label{sec:orders}

We continue to substitute \(\eta_{0\alpha}\) with \Eq{eq:eta} in order to further write \(H_{\text{QZD}}\) into a power series
\al{
H_{\text{QZD}}
&=\lambda^{-1}\sum_{\alpha}\cancelto{0}{{\eta}^{(0)}_{0}}P^{(0)}_{0\alpha}\nonumber\\&+\sum_{\alpha}{\eta}^{(1)}_{0\alpha}P^{(0)}_{0\alpha}+\lambda\sum_{\alpha}{\eta}^{(2)}_{0\alpha}P^{(0)}_{0\alpha}+\mathcal{O}(\lambda^2)\nonumber\\
&\equiv H^{(0)}_{\text{QZD}}+\lambda H^{(1)}_{\text{QZD}}+\mathcal{O}(\lambda^2). 
\label{eq:hqzd}
}
The order of coherent QZD is determined by the order of the largest term in the above series expansion of \(H_{\text{QZD}}\), which does not commute with \(\ket{\psi(0)}\bra{\psi(0)}\). The term will be the largest contributor to the constrained dynamics in coherent QZD.

\subsection{Zeroth-order coherent QZD}

\begin{figure}[htp!]
  \centering
  \includegraphics[width=0.75\linewidth]{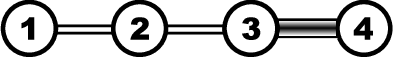}
   \caption{Schematics of a 4-level system, where we represent the levels with circles placed horizontally. The narrow and broad lines represent weak coupling \(k\) and strong coupling \(\lambda ^{-1}k\) respectively.}
\label{fig:chain0}
\end{figure}
The zeroth-order coherent QZD, i.e., \(H^{(0)}_{\text{QZD}}\) contributes the most, occurs when the degeneracy in \({\eta}^{(0)}_{0}\) is lifted to first order. According to degenerate perturbation theory, in this case, \(\ket{\phi^{(0)}_{0\alpha}}\) should diagonalize \(P^{(0)}_0HP^{(0)}_0\), so that
\begin{align}
\brak{\phi^{(0)}_{0\alpha}}{H}{\phi^{(0)}_{0\beta}}=\eta^{(1)}_{0\alpha}\delta_{\alpha\beta},
\label{eq:firstdiag}
\end{align}
and \(\eta^{(1)}_{0\alpha}\neq\eta^{(1)}_{0\beta}\), whenever \(\alpha\neq\beta\). This renders \al{H^{(0)}_{\text{QZD}}\equiv\sum_{\alpha}\eta^{(1)}_{0\alpha}P^{(0)}_{0\alpha}=\sum_{\alpha}P^{(0)}_{0\alpha}HP^{(0)}_{0\alpha}=P^{(0)}_0HP^{(0)}_0.
\label{eq:h0_0}
}
\([H^{(0)}_{\text{QZD}},\ket{\psi(0)}\bra{\psi(0)}]\neq0\) unless \(\ket{\psi(0)}\) coincides one of the eigenstates \(\ket{\phi^{(0)}_{0\alpha}}\). (If so, all terms in \Eq{eq:hqzd} will commute with \(\ket{\psi(0)}\bra{\psi(0)}\) and there will be no dynamics at all.) Therefore, the largest contributor to QZD in \Eq{eq:hqzd} should be \(H^{(0)}_{\text{QZD}}=P^{(0)}_0HP^{(0)}_0\). This case is called the zeroth-order coherent QZD.

The simplest example of the zeroth-order coherent QZD is given by Facchi \etal \cite{facchiQuantumZenoSubspaces2002,facchiQuantumZenoSubspaces2003}, a 4-level system illustrated by \Figure{fig:chain0}, where \(\ket{\psi(0)}=\ket{1}\), \(H=k\ket{1}\bra{2}+k\ket{2}\bra{3}+\textit{h.c.}\), \( \lambda^{-1}H_w=\lambda^{-1}k\ket{3}\bra{4}+\textit{h.c.}\) and \(P^{(0)}_0=\ket{1}\bra{1}+\ket{2}\bra{2}\). According to \Eq{eq:timeop}, we derive
\al{
\lim_{{\lambda}\to0}e^{-iH_\text{tot}t}\ket{\psi(0)}=e^{-iH^{(0)}_{\text{QZD}}t}\ket{\psi(0)},
\label{eq:lim}
}
where \(H^{(0)}_{\text{QZD}}=P^{(0)}_0HP^{(0)}_0=k\ket{1}\bra{2}+\textit{h.c.}\) This means that the strong coupling between 3 and 4 counter-intuitively traps the dynamics within a subspace merely containing 1 and 2. 

In fact, the experimentally realized coherent QZD mentioned earlier all belong to the zeroth order \cite{schaferExperimentalRealizationQuantum2014,signolesConfinedQuantumZeno2014,bretheauQuantumDynamicsElectromagnetic2015,linPreparationEntangledStates2016}. The effect of \(P^{(0)}_0HP^{(0)}_0\) is regarded as a resource to engineer the Hilbert space and to manipulate the quantum state \cite{burgarth2014exponential,bretheauQuantumDynamicsElectromagnetic2015,linPreparationEntangledStates2016}. It has multiple potential applications, such as generating entangled states \cite{barontiniDeterministicGenerationMultiparticle2015,linPreparationEntangledStates2016} or other useful states \cite{touzardCoherentOscillationsQuantum2018,bretheauQuantumDynamicsElectromagnetic2015,raimond2010phase}, such as squeezed states of light. In contrast, the first-order coherent QZD (FC-QZD) has barely been investigated, the reason for which will be explained after its introduction.

\subsection{First-order coherent QZD}

FC-QZD, i.e., \(\lambda H^{(1)}_{\text{QZD}}\) contributes the most, occurs when the degeneracy in \({\eta}^{(0)}_{0}\) is NOT lifted to first order, but to second order. In this case, \Eq{eq:firstdiag} remains true but \(\eta^{(1)}_{0\alpha}=\eta^{(1)}_{0\beta}=\dots=\eta^{(1)}_{0}\). This entails \(H^{(0)}_{\text{QZD}}=\eta^{(1)}_{0}P^{(0)}_0\). Recalling \(\ket{\psi(0)}\in P^{(0)}_{0}\mathcal{H}_\text{tot}\), we obtain \([\eta^{(1)}_{0}P^{(0)}_0,\ket{\psi(0)}\bra{\psi(0)}]=0\). This suggests that \(H_{\text{QZD}}^{(0)}\) no longer amounts to any dynamics. 

In this case, \(\ket{\phi_{0\alpha}^{(0)}}\) must also diagonalize \(P^{(0)}_{0}H{\widetilde{Q}^{(0)}_0}HP^{(0)}_{0}\), so that
\begin{align}
\brak{\phi_{0\alpha}^{(0)}}{H{\widetilde{Q}^{(0)}_0}H}{\phi_{0\beta}^{(0)}}=\eta^{(2)}_{0\alpha}\delta_{\alpha\beta}~\text{with}~{\widetilde{Q}^{(0)}_0}\equiv \sum_{n\neq 0} \dfrac{{P^{(0)}_{n}}}{-\eta^{(0)}_n},
\label{eq:seconddiag}
\end{align}
and \(\eta^{(2)}_{0\alpha}\neq\eta^{(2)}_{0\beta}\) whenever \(\alpha\neq\beta\). This yields
\al{
\lambda H^{(1)}_{\text{QZD}}\equiv\lambda\sum_{\alpha}\eta^{(2)}_{0\alpha}P^{(0)}_{0\alpha}&=\lambda\sum_{\alpha}P^{(0)}_{0\alpha}H{\widetilde{Q}^{(0)}_0}HP^{(0)}_{0\alpha}\nonumber\\&=\lambda P^{(0)}_{0}H{\widetilde{Q}^{(0)}_0}HP^{(0)}_{0}
\label{eq:hqzd1}.
}
Once \([\lambda H^{(1)}_{\text{QZD}},\ket{\psi(0)}\bra{\psi(0)}]\neq 0\) is confirmed, the largest contributor to QZD should be \(\lambda H^{(1)}_{\text{QZD}}\). This case is called the \textit{first-order coherent QZD} (FC-QZD). 

Because \(\lambda H^{(1)}_{\text{QZD}}\) only exists when \(\lambda\) is non-zero, \(\mathcal{O}(\lambda)\ket{\psi(0)}\) in \Eq{eq:timeop} should also be non-zero, leading to a population leakage out of \(P_0^{(0)}\mathcal{H}_\text{tot}\). Consequently, realizing FC-QZD through \(e^{-iH_\text{tot}t}\ket{\psi(0)}\) requires concession by allowing moderate leakage. Therefore, we denote the largest ever leakage  by \(\delta\), namely the largest ever population in \((1-P_0^{(0)})\mathcal{H}_\text{tot}\) throughout the interested period. The standard of ``moderate" can be quantified as \(\delta<\delta_0\), where \(\delta_0\) is a standard to be determined according to needs. 

To summarize, the following are the key prerequisites for FC-QZD:
\begin{enumerate}
  \item[\textbf{(I)}] \(H^{(0)}_{\text{QZD}}\propto P^{(0)}_{0}\), \([\lambda H^{(1)}_{\text{QZD}},\ket{\psi(0)}\bra{\psi(0)}]\neq 0\);
  \item[\textbf{(II)}] \(\delta<\delta_0\).
\end{enumerate}
From the above, it is easy to infer the criteria for coherent QZD at higher orders, e.g., the second order should require \(H^{(0)}_{\text{QZD}},~H^{(1)}_{\text{QZD}}\propto P^{(0)}_{0}\) and \([\lambda^2 H^{(2)}_{\text{QZD}},\ket{\psi(0)}\bra{\psi(0)}]\neq 0\).

FC-QZD has received little attention because it was not included in the original definition of coherent QZD. When Facchi \etal made the original definition \cite{facchiQuantumZenoSubspaces2002}, they focused on the extreme case \(\lambda \to0\) which buries the existence of FC-QZD since \(\lambda H^{(1)}_{\text{QZD}}\) vanishes, let alone higher orders. Hence, the original definition of coherent QZD merely recognizes the zeroth-order type since only \(H^{(0)}_{\text{QZD}}\) survives when \(\lambda \to0\) as we saw in the example \Figure{fig:chain0}. Thus, it is no surprise that the zeroth-order coherent QZD is followed by a considerable number of experiments and investigations mentioned before, whereas the other orders are not noticed. However, a finite \(\lambda\) is what one can actually achieve in real life, so FC-QZD is rather practical and should be recognized.

The same problem also occurs for another type of QZD realized by measurements. Although Facchi \etal applied infinitely frequent measurements to define QZD \cite{facchiQuantumZenoSubspaces2002}, latter studies have found that it has a rank of orders considering a finite frequency \cite{elliottQuantumQuasiZenoDynamics2016,Dhar2015}. Those orders are in parallel with ours. For FC-QZD, its counterpart is called second-order quantum quasi-Zeno dynamics (QqZD), which has proven capable of inducing a long-range correlated exchange in many-body systems like spin chains \cite{elliottQuantumQuasiZenoDynamics2016,kozlowskiNonHermitianDynamicsQuantum2016,mazzucchiQuantumMeasurementinducedDynamics2016}. The second-order QqZD is described by an effective Hamiltonian of a similar form as \(H_{\text{QZD}}^{(1)}=P^{(0)}_{0}H{\widetilde{Q}^{(0)}_0}HP^{(0)}_{0}\). Due to the similarity, we infer that FC-QZD can also generate a non-local transition, which is confirmed below. 

\section{FC-QZD in tight-binding chains}
\label{sec:chain}
\begin{figure}[h]
  \centering
  \includegraphics[width=0.9\linewidth]{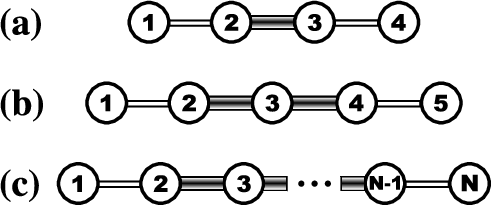}
  \caption{Schematics of tight-binding chains with different lengths. The narrow and broad lines connecting the nearest sites represent weak coupling \(k\) and strong coupling \(\lambda^{-1}k\) respectively.}
\label{fig:chain}
\end{figure}

Quantum systems like \Figure{fig:chain0} with consecutive couplings only between the neighbored levels are called tight-binding chains by us. These chains are subject to a tight-binding Hamiltonian generally written as \(H_{tb}=\sum_{i}k_i \smlb{\ket{i}\bra{i+1}+h.c.}\). Tight-binding model is widely used in different fields, such as solid-state physics \cite{feynman1965feynman,landinote}, spin chains \cite{bose2007quantum,Murphy2010,kiely2021fast}, and light propagation in waveguides \cite{longhi2006,liuEngineeringZenoDynamics2023,Chen2021}.

Next, we construct some special chains as illustrated by \Figure{fig:chain} with lengths \(N\geq4\), where the weak interactions are placed only at the two ends. The resulting Hamiltonian reads:
\al{
H_\text{tot}&=\lambda^{-1} H_w+H \nonumber\\
&
\begin{aligned}
    &=\lambda^{-1} k\midb{\sum_{i=2}^{N-2} \ket{i}\bra{i+1}+\textit{h.c.}}\\
&\qquad\qquad+k\Big[\big(\ket{1}\bra{2}+\ket{N-1}\bra{N}\big)+\textit{h.c.}\Big],
\end{aligned}
\label{eq:hChain}
} 
where \(\lambda^{-1} k\) and \({k}\) are the strong and the weak coupling constants, respectively. Suppose that the initial state is \(\ket{\psi(0)}=\ket{1}\) so that \(H_w\ket{1}=0\). Surprisingly, we find that chains with an even \(N\) are capable of realizing FC-QZD while chains with an odd \(N\) are not. In the following, we prove this claim by checking the prerequisites (I, II). 

\subsection{Even chain}

\begin{figure*}[ht]
  \centering
\includegraphics[width=\linewidth]{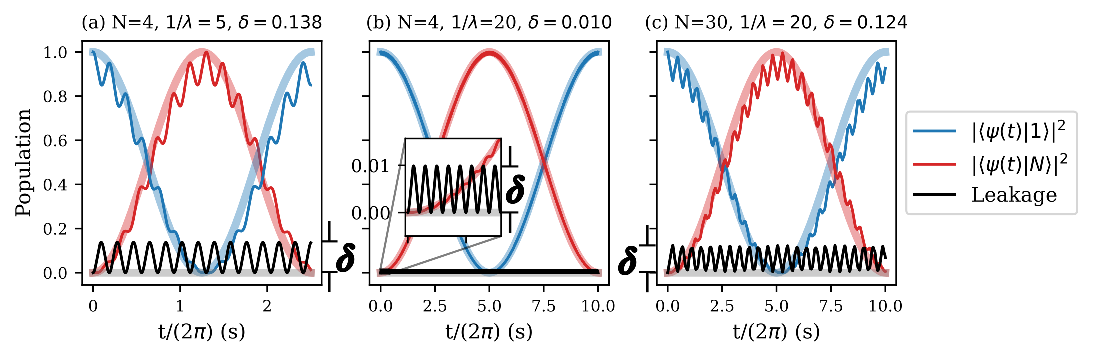}
  \caption{Population evolution on even chains. Set \(k=1\) thereafter. The narrow solid curves depict the prediction given by \(\ket{\psi(t)}=e^{-iH_\text{tot}t}\ket{1}\) where \(H_{\text{tot}}\) is defined in \Eq{eq:hChain}, of the populations at two ends (blue and red) and the leakage (black), which is the population outside the two ends. The broad curves in the background illustrate the prediction of \(\lambda H_{\text{QZD}}^{(1)}\) \Eq{eq:4chain}. (a) 4-site chain with \(\lambda^{-1}=5\) has a leakage \(\delta\) beyond the upper bound \(\delta_0=0.1\). (b) We increase \(\lambda^{-1}=20\) and the leakage is made below 0.1 shown in the inset. This dynamic reaches our standard for FC-QZD manifesting itself as a non-local end-to-end transition. (c) The chain is extended to \(N=30\) and again obtains a leakage over 0.1.}
\label{fig:popu_even}
\end{figure*}
\label{sec:even}
\subsubsection{it cannot be zeroth-order}
When N is even, \(H_w\) has eigenprojector \(P^{(0)}_0=\ket{1}\bra{1}+\ket{N}\bra{N}\) (see \App{sec:evenH2} for its proof). Knowing that \(H\) only connects \(\ket{1}\) to \(\ket{2}\) and \(\ket{N-1}\) to \(\ket{N}\), we get \(H_{\text{QZD}}^{(0)}=P^{(0)}_0HP^{(0)}_0=0\). We conclude that it is not the zeroth-order but probably FC-QZD.

\subsubsection{FC-QZD: a non-local transition between two ends}
We continue to solve \(\lambda H_{\text{QZD}}^{(1)}\) in the shortest even chain owning 4 sites in \Figure{fig:chain}(a). The non-zero eigenvalues and the corresponding eigenstates of \(H_w\) are 
\begin{align}
\begin{aligned}
\eta_{+}=k,&\quad \ket{\phi_+}=(\ket{2}+\ket{3})/\sqrt{2},\\
\eta_{-}=-k,&\quad \ket{\phi_-}=(\ket{2}-\ket{3})/\sqrt{2}.
\end{aligned}
\end{align}
Plugging \(P^{(0)}_\pm=\ket{\phi_\pm}\bra{\phi_\pm}\) into \Eq{eq:hqzd1}, we obtain
\al{
\lambda H_{\text{QZD}}^{(1)}&=\lambda P^{(0)}_{0}H\frac{P^{(0)}_+}{-\eta_{+}}HP^{(0)}_{0}+\lambda P^{(0)}_{0}H\frac{P^{(0)}_-}{-\eta_{-}}HP^{(0)}_{0}\nonumber\\
&=-\lambda k\smlb{\ket{1}\bra{4}+\ket{4}\bra{1}}
\label{eq:4chain}
}
directly connecting the two ends, which disobeys the tight-binding model only binding the nearest neighbors. In other words, \(\lambda H_{\text{QZD}}^{(1)}\) describes a \textit{non-local end-to-end} population transition, reminiscent of Newton's cradle. Moreover, we fulfill (I) since \(\midb{\lambda H^{(1)}_{\text{QZD}},\ket{1}\bra{1}}\neq 0\). 

However, according to (II), the dynamics \(e^{-iH_\text{tot}t}\ket{1}\) can be recognized as FC-QZD, only if the leakage \(\delta<\delta_0\). In this case, \(P^{(0)}_{0}\mathcal{H}_\text{tot}\) refers to two ends and, therefore, the leakage is the population outside the two ends. According to \Eq{eq:timeop}, the leakage can be suppressed by increasing \(\lambda^{-1}\). Then we want to test whether \({\lambda^{-1}}=5\) is sufficient, so we plot the dynamics \(\ket{\psi(t)}=e^{-iH_\text{tot}t}\ket{1}\) in \Figure{fig:popu_even}(a) with narrow solid curves. In the background, the broad lines illustrate the prediction by \(\lambda H^{(1)}_{\text{QZD}}\) in \Eq{eq:4chain}. By comparison, \(H_\text{tot}\) overall gives a similar pattern to \(\lambda H^{(1)}_{\text{QZD}}\), that the population oscillates between the left end (blue) and the right end (red). But a small part of the population periodically leaks out of two ends (black). \(\delta\) is defined as the largest leakage, i.e. the maximum height of the black oscillating curve. Say \(\delta_0=0.1\), the 4-site chain with \(\lambda^{-1}=5\) renders \(\delta=0.138>\delta_0\), which fails (II). In other words, \(\lambda^{-1}=5\) is not large enough.

Next, we increase \({\lambda^{-1}}=20\) and record the new dynamics in \Figure{fig:popu_even}(b) to see whether it is sufficient. At first glance, the predictions given by \(\lambda H_{\text{QZD}}^{(1)}\) and \(H_\text{tot}\) almost overlap now, being a non-local end-to-end transition. However, the inset highlights the minor leakage induced by \(H_\text{tot}\). As marked in the inset, \(\lambda^{-1}=20\) renders \(\delta=0.01<0.1\), which successfully checks (II). Eventually, we can say that FC-QZD is realized in a chain with length \(N=4\) and \(\lambda^{-1}=20\). However, the truth of non-local transition is that, population first leaves \(\ket{1}\in P^{(0)}_0\mathcal{H}_\text{tot}\), bypasses intermediate \((1-P^{(0)}_0)\mathcal{H}_\text{tot}\) at a quick enough rate and arrives \(\ket{4}\in P^{(0)}_0\mathcal{H}_\text{tot}\), which reminds us of Raman scattering with a virtual energy level.   

For even chains with an arbitrary length N like \Figure{fig:chain}(c), we derive a general expression (see \App{sec:evenH2} for derivation):
\al{
\lambda H_{\text{QZD}}^{(1)}(\text{even}~N)=(-1)^{\frac{N}{2}-1}\lambda k\smlb{\ket{1}\bra{N}+h.c.}.
\label{eq:evenN}
}
It seems that FC-QZD in the manner of an end-to-end transition can occur on all even chains at a rate \(\lambda k\). However, it should be impossible for the end-to-end transition across an infinitely long chain to consume only a finite time \({\lambda^{-1}\pi/k}\), unless \(\lambda^{-1}\to\infty\). 

To test this conjecture, the evolution of a 30-site even chain with \({\lambda^{-1}}=20\) is recorded in \Figure{fig:popu_even}(c). Evidently, the large leakage \(\delta>0.1\) returns. It means \({\lambda^{-1}}=20\) is no longer sufficient for the 30-site chain. This raises another question: how large \({\lambda^{-1}}\) should be to satisfy (II) on an even chain with an arbitrary length N. In \App{sec:validity}, we derive its answer for \(\delta_0<0.2\) that one needs
\al{
\lambda^{-1}>{f(N)}{\sqrt{4.3/\delta_0}},
}
where \(f(N)\equiv\tan{\smlb{\frac{\pi}{2}\frac{N-2}{N-1}}}/{\sqrt{N-1}}\) is a monotonically increasing function of N. This implies that for an infinitely long chain, if FC-QZD is realized, \({\lambda^{-1}}\) should be infinite and thus the end-to-end transition should have an infinitely long period.

In addition, \Figure{fig:popu_even} exhibits that, the leakage always experiences an increase in its amplitude and a decrease in its frequency at the same time, for which we will also provide a reason at the end of \App{sec:validity}.

\subsection{Odd chain}

\subsubsection{FC-QZD is no longer available}
In short, adding one single site to an even chain surprisingly will ruin FC-QZD. \(H_w\) with an odd N has \(P_0^{(0)}=\ket{1}\bra{1}+\ket{N}\bra{N}+\ket{\phi_{0\text{mid}}}\bra{\phi_{0\text{mid}}}\) (see \App{sec:odd_detune} for details), where
\al{
\ket{\phi_{0\text{mid}}}=\dfrac{1}{\sqrt{(N-1)/2}}\midb{\ket{2}-\ket{4}+...+(-1)^{\frac{N+1}{2}}\ket{N-1}}.
\label{eq:phi_mid}
}
Consequently, the existence of \(\ket{\phi_{0\text{mid}}}\) revives \(H_{\text{QZD}}^{(0)}\):
\al{
H_{\text{QZD}}^{(0)}&(\text{odd}~N)=P_0^{(0)}HP_0^{(0)}\nonumber\\&=\frac{k}{\sqrt{(N-1)/2}}\smlb{\ket{1}\bra{\phi_{0\text{mid}}}+\ket{\phi_{0\text{mid}}}\bra{N}+h.c.},
\label{eq:odd1}
}
so \([H^{(0)}_{\text{QZD}},\ket{\psi(0)}\bra{\psi(0)}]\neq 0\). Therefore, it is NOT FC-QZD, but zeroth-order. 

Unlike \(\lambda H_{\text{QZD}}^{(1)}\) in \Eq{eq:evenN}, \(H_{\text{QZD}}^{(0)}\) is not an end-to-end transition due to the existence of the intermediate state \(\ket{\phi_{0\text{mid}}}\). Again to give the simplest example, the dynamics of the 5-site chain is depicted in \Figure{fig:popu_odd}(a). The green curve shows that \(\ket{\phi^{N=5}_{0\text{mid}}}=(\ket{2}-\ket{4})/\sqrt{2}\) periodically acquires up to half of the entire population.

\subsubsection{How to revive FC-QZD}
\label{sec:odd_2}

\begin{figure}[ht!]
  \centering
  \includegraphics[width=\linewidth]{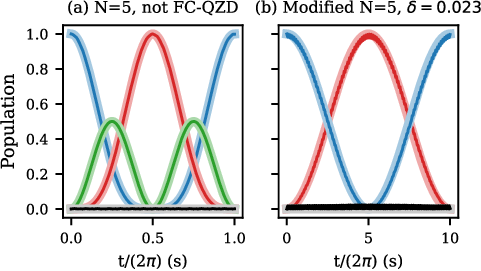}
  \caption{(a) An odd 5-site chain manifests zeroth-order coherent QZD. The broad curves in the background now represent the prediction given by \(H_{\text{QZD}}^{(0)}\) in \Eq{eq:odd1} whereas the narrow curves still illustrate \(e^{-iH_\text{tot}t}\ket{1}\). We omit the legend because the meanings of the colors are the same as \Figure{fig:popu_even}, except for the green curve illustrating \(\abs{\braket{\psi(t)}{\phi_{0\text{mid}}}}^2\). (b) After modification \Eq{eq:hmd} with \(\Delta\omega=\lambda^{-1}k\), the odd 5-site chain with \(\lambda^{-1}=20\) retrieves FC-QZD in the form of an end-to-end transition. Now, the narrow curves correspond to \(H_\text{mod}\) in \Eq{eq:hmd} and the broad curves in the background correspond to \(\lambda H_{\text{QZD}}^{(1)}(\text{odd}~N)\) in \Eq{eq:odd_mod}.}
\label{fig:popu_odd}
\end{figure}
To retrieve FC-QZD in odd chains, a possible solution is to introduce a non-zero value \(\Delta\omega\) at the 2nd diagonal element of \(H_{\text{tot}}\), written as 
\al{
H_\text{mod}=H_{\text{tot}}+\Delta\omega\ket{2}\bra{2}.
\label{eq:hmd}
}
As shown in \App{sec:odd_detune}, \(H_\text{mod}\) retrieves \(P_0^{(0)}=\ket{1}\bra{1}+\ket{N}\bra{N}\) and then forces \(H_{\text{QZD}}^{(0)}=0\), reviving FC-QZD. Note that the diagonal elements of \(H_{\text{tot}}\) are all zero because all sites are supposed to have identical on-site energy. So \(\Delta\omega\ket{2}\bra{2}\) refers to a shift in energy on site 2. 

For a modified odd chain with a length N, we derive (see \App{sec:odd_detune} for details)
\al{
\lambda H_{\text{QZD}}^{(1)}(\text{odd}~N)=&(-1)^{\frac{N-1}{2}}\dfrac{k^2}{\Delta\omega}\smlb{\ket{1}\bra{N}+h.c.}\nonumber\\&-\frac{{k}^2}{\Delta\omega}\smlb{\ket{1}\bra{1}+\ket{N}\bra{N}}.
\label{eq:odd_mod}
}
We take the modified 5-site chain with \(\Delta\omega=\lambda^{-1}k\) and \(\lambda^{-1}=20\) as an example and illustrate its dynamics \(e^{-iH_\text{mod}t}\ket{1}\) in \Figure{fig:popu_odd}(b). With \(\delta=0.023<0.1\), FC-QZD in the form of an end-to-end transition is successfully revived in the modified 5-site chain. Odd chains should also suffer from increasing length and the value of \(\Delta\omega\) should not be arbitrary, but we defer the discussion to future research.

\section{Discussion}

It is interesting that the constraints in coherent QZD seem to have a principle of action regardless of orders: to preserve \(\brak{\psi(t)}{\lambda^{-1} H_w}{\psi(t)}\) which used to be zero at \(t=0\). This can be inferred from the observation that a smaller \(\lambda^{-1}\) and a larger \(N\) results in a larger \(\delta\). According to \Eq{eq:etan}, either decreasing \(\lambda^{-1}\) or increasing N decreases the spacing between zero level and its neighbors in the spectrum of \(\lambda^{-1} H_w\). Thus, it is possible for the quantum state to leave \(P_0^{(0)}\mathcal{H}_\text{tot}\) corresponding to zero energy for somewhere else without changing too much in \(\avg{\lambda^{-1} H_w}\). This indicates a weaker constraint or a larger leakage. Besides, dynamics inside \(P_0^{(0)}\mathcal{H}_\text{tot}\) is always allowed since during which \(\avg{\lambda^{-1} H_w}=0\) is always preserved. 

To experimentally implement FC-QZD on tight-binding chains, we think the optical waveguide array used in \cite{liuEngineeringZenoDynamics2023} is a good platform. The propagation of light in the waveguide array is described by tight-binding model, where its wave equation is parallel to the Schr\"{o}dinger equation. In principle, the individual coupling intensity can be adjusted by adjusting the spacing between the waveguides. A direct analog for on-site energy is the propagation coefficient, which can be tuned by the refractive indices of waveguides. Not to mention, a structure quite similar to the 4-site chain has already appeared in Figure 3 of \cite{liuEngineeringZenoDynamics2023}. Spin chains and real 1-D lattices are surely microscopic tight-binding systems governed by quantum mechanics. But they are much harder to manipulate, with less flexibility for parameter adjustment. At the end of \App{sec:evenH2}, we show that FC-QZD is robust against fluctuation in coupling intensities, namely \(k\), proving its feasibility in real life.

As mentioned before, previous researchers have found that dynamics under \(H^{(0)}_{\text{QZD}}=P_0^{(0)}{H}P_0^{(0)}\) constitutes a resource for the quantum state manipulation through its ability to constrain the accessible regime in Hilbert space \cite{burgarth2014exponential,barontiniDeterministicGenerationMultiparticle2015,linPreparationEntangledStates2016,touzardCoherentOscillationsQuantum2018,bretheauQuantumDynamicsElectromagnetic2015,raimond2010phase}. FC-QZD with \(H^{(1)}_{\text{QZD}}=P^{(0)}_{0}H{\widetilde{Q}^{(0)}_0}HP^{(0)}_{0}\) provides another type of constraint, adding another piece to the toolbox for quantum manipulation. Considering the non-local transition seen in the chains, FC-QZD is promising to create more interesting quantum behavior. 
\begin{acknowledgments}
I would like to express my sincere gratitude to Prof. Joseph H. Eberly for his invaluable guidance and support throughout the course of this project.
\end{acknowledgments}

\appendix

\section{How to guarantee (II) \(\delta<\delta_0\) for an even chain with a length N by adjusting \(\lambda\)?}
\label{sec:validity}

First, we want to prove that \(\mathcal{O}(\lambda)\ket{\psi(0)}\) in \Eq{eq:timeop} induces the leakage. From \Eq{eq:uprime}, 
\al{\mathcal{O}(\lambda)=\sum_{n\alpha}\text{exp}\smlb{-i{\eta}_{n\alpha}\tau}\smlb{ \lambda P^{(1)}_{n\alpha}+\mathcal{O}(\lambda^2)},}
where \(\tau\equiv t/\lambda\). Its largest contributor is denoted
\al{
U^{(1)}&(\tau)\equiv \lambda\sum_{n\alpha}\text{exp}\smlb{-i{\eta}_{n\alpha}\tau}P^{(1)}_{n\alpha}\nonumber\\&=\lambda\sum_{n\alpha}\text{exp}\smlb{-i{\eta}_{n\alpha}\tau}\smlb{\ket{\phi_{n\alpha}^{(1)}}\bra{\phi_{n\alpha}^{(0)}}+\ket{\phi_{n\alpha}^{(0)}}\bra{\phi_{n\alpha}^{(1)}}}.
\label{eq:o2}
}
From the equation above, we know \(U^{(1)}(\tau)\) can at least transfer \(\ket{\phi_{0\alpha}^{(0)}}\in P_0^{(0)}\mathcal{H}_\text{tot}\) into \(\ket{\phi_{0\alpha}^{(1)}}\). Reminding the well-known conclusion about the perturbed eigenstates, whose first-order correction, namely \(\ket{\phi_{0\alpha}^{(1)}}\), always contains unperturbed eigenstates of the rest energy levels, namely \(\ket{\phi_{n\neq0}^{(0)}}\notin P_0^{(0)}\mathcal{H}_\text{tot}\). Therefore, \(\ket{\phi_{0\alpha}^{(1)}}\) is not in the subspace \(P_0^{(0)}\mathcal{H}_\text{tot}\) either. Eventually, we conclude that \(U^{(1)}\) and thus \(\mathcal{O}(\lambda)\) must transfer population from the subspace to the outside, namely causing leakage.

In the context of even tight-binding chains, we can derive the expression of \(\abs{U^{(1)}(\tau)\ket{\psi(0)}}^2\) in terms of \(\lambda\) and \(N\). According to the analysis above, the maximum of \(\abs{U^{(1)}(\tau)\ket{\psi(0)}}^2\) should provide an estimation of \(\delta\), which is also in terms of \(\lambda\) and \(N\). This relation will answer how to guarantee (II) \(\delta<\delta_0\) for any length N by adjusting \(\lambda^{-1}\).

To derive \(\abs{U^{(1)}\ket{\psi(0)}}^2\), we first solve the unperturbed degenerate states with respect to \(\eta^{(0)}_0=0\) by diagonalizing the Hamiltonian \(H_{\text{QZD}}^{(1)}\) in \Eq{eq:evenN}:
\al{
\begin{aligned}
\ket{\phi^{(0)}_{0\alpha}}=\dfrac{\ket{1}+\ket{N}}{\sqrt{2}},~
\ket{\phi^{(0)}_{0\beta}}=\dfrac{\ket{1}-\ket{N}}{\sqrt{2}}.
\end{aligned}
\label{eq:diag}
}
According to a mathematical work about the tridiagonal Toeplitz matrix \cite{noschese2013tridiagonal}, the rest eigenvalue of \(H_w\) is given by 
\al{
\eta^{(0)}_n=2k\cos{\smlb{\frac{n\pi}{N-1}}},~n=1,\dots,N-2,
\label{eq:etan}
}
where \(-2k<{\eta^{(0)}_n}<2k\). The corresponding eigenvector is \(\ket{\phi_{n}^{(0)}}=\sqrt{\frac{2}{N-1}}\midb{0,x_{n2},x_{n3},\dots,x_{n(N-1)},0}^T\), where
\al{
x_{ni}=\sin{\smlb{\dfrac{n(i-1)\pi}{N-1}}},~i=2,3,\dots,N-1.
}
Meanwhile, the first-order correction to eigenstates are
\al{
\begin{aligned}
&\ket{\phi_{n\neq 0}^{(1)}}=\sum_{i\neq n~\text{or}~0} \ket{\phi^{(0)}_i}\dfrac{H_{i,n}}{\eta^{(0)}_n-\eta^{(0)}_{i}}\\&\qquad\qquad+\ket{\phi^{(0)}_{0\alpha}}\dfrac{H_{0\alpha,n}}{\eta^{(0)}_n}+\ket{\phi^{(0)}_{0\beta}}\dfrac{H_{0\beta,n}}{\eta^{(0)}_n},\\
&\ket{\phi_{0\alpha(\beta)}^{(1)}}=\sum_{n\neq 0} \ket{\phi^{(0)}_n}\dfrac{H_{n,0\alpha(\beta)}}{-\eta^{(0)}_{n}},
\end{aligned}
\label{eq:first_phi}
}
where we remove the subscripts of degeneracy for \(n\neq0\) because only \(n=0\) has degeneracy in this chain model (verified by \Eq{eq:etan}). For simplicity, the matrix element \(H_{m,n}\) is abbreviation for \(\brak{\phi_{m}^{(0)}}{H}{\phi_{n}^{(0)}}\).  

\begin{figure*}[ht!]
  \centering
  \includegraphics[width=\linewidth]{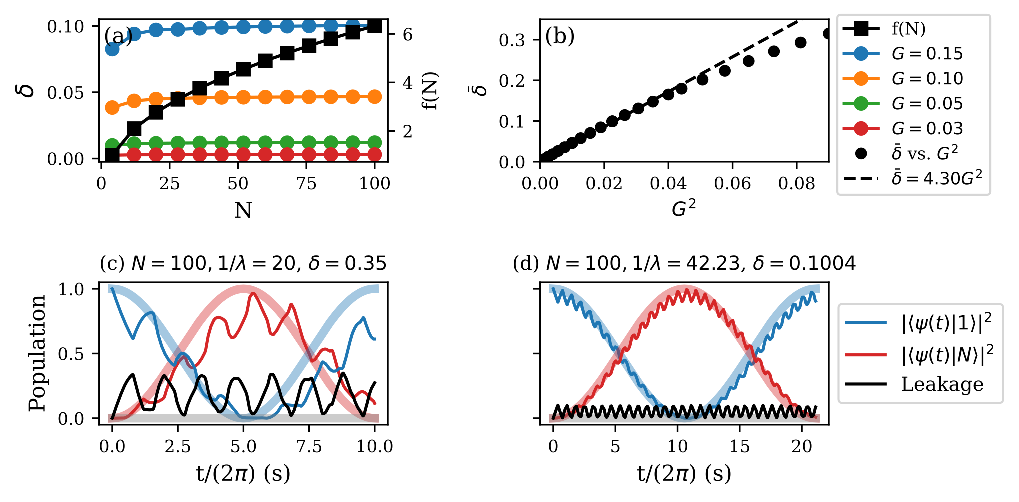}
  \caption{(a) For each line with circles, we assign a value for \(G\). In each point, \(\delta\) is calculated from \(H_\text{tot}\) in \Eq{eq:hChain} with corresponding \(N\) (x coordinate) and \({\lambda}= G/f(N)\) inserted. The lines are almost flat, with an average height denoted \(\bar{\delta}\). Clearly, \(\bar{\delta}\) increases with \(G\). The line with squares plots \(f(N)\) with respect to the right y-axis. (b) By keeping varying \(G\), we collect more resulting \(\bar{\delta}\) to observe their numerical relationship. For \(\bar{\delta}<0.2\), \(\bar{\delta}\) is almost proportional to \(G^2\) where fitting finds its slope equal to 4.30. (c) For a 100-site chain, \({\lambda^{-1}}=20\) renders \(\delta>0.1\). (d) By increasing \({\lambda^{-1}}=42.23\), \(\delta\) reaches the upper bound 0.1.}
\label{fig:popu_acc}
\end{figure*}

Insert above into \Eq{eq:o2} and act on \(\ket{\psi(0)}=\ket{1}\). We obtain
\al{
&U^{(1)}(\tau)\ket{1}=\sum_{n=1,3,\dots,N-3}\Bigg[\ket{\phi_{n}^{(0)}}\smlb{e^{-i{\eta}_{n}\tau}-e^{-i{\eta}_{0\alpha}\tau}}\dfrac{g_n}{\sqrt{2}}\Bigg]\nonumber\\
&+\sum_{n=2,4,\dots,N-2}\Bigg[\ket{\phi_{n}^{(0)}}\smlb{e^{-i{\eta}_{n}\tau}-e^{-i{\eta}_{0\beta}\tau}}\dfrac{g_n}{\sqrt{2}}\Bigg],
\label{eq:ut1_after_plugin_matrix_ele}
}
where we define
\al{
g_n(N,{\lambda})\equiv\dfrac{\lambda}{\sqrt{N-1}}\tan{\smlb{\dfrac{n\pi}{N-1}}}.
\label{eq:gn}
}  
It is not difficult to find that the maximum \(\abs{g_n}\) is at \(n=N/2-1\) or \(n=N/2\), which is denoted by
\al{
G(N,{\lambda})\equiv\lambda f(N),
\label{eq:analy}
} 
where \(f(N)\equiv\tan{\smlb{\frac{\pi}{2}\frac{N-2}{N-1}}}/{\sqrt{N-1}}\) is a monotonically increasing function as illustrated by the black square line \Figure{fig:popu_acc}(a). According to analysis before, because the maximum of \(\abs{U^{(1)}(\tau)\ket{1}}^2\sim G^2\), \(G^2\) should form a good estimation for \(\delta\).

To demonstrate inference, we plot \Figure{fig:popu_acc}(a) and (b). \Figure{fig:popu_acc}(a) proves that, although \(N\) and \(\lambda\) varies, resulting \(\delta\) is almost invariant with respect to an average \(\bar{\delta}\), as long as \(G(N,\lambda)\) is preserved. Also, we observe that \(\bar{\delta}\) increases as \(G\) increases. \Figure{fig:popu_acc}(b) includes more combinations of parameters and shows that \(\bar{\delta}\) almost grows linearly with \(G^2\) within the regime \(\bar{\delta}<0.2\). Fitting yields \al{
\delta\approx\bar{\delta}=4.3G^2(N,{\lambda}),
\label{eq:dg}
}
which is the analytical estimation of \(\delta\) in terms of \(N\) and \(\lambda\) we pursue.

Therefore, for an even chain with length \(N\), we can guarantee \(\delta<\delta_0\) through the restriction
\als{
\lambda^{-1}>{f(N)}{\sqrt{4.3/\delta_0}}
}
for any standard \(\delta_0<0.2\). Taken \(\delta_0=0.1\) and \(N=100\) as an example, \(f(100)\sqrt{4.3/0.1}=42.23\). In \Figure{fig:popu_acc}(c), \({\lambda^{-1}}=20<42.23\) does not meet (II). When we adjust \({\lambda^{-1}}=42.23\), \(\delta\) arrives right on the boundary \(0.1\) as shown in \Figure{fig:popu_acc}(d). Thus, any \({\lambda^{-1}}>42.23\) will definitely retrieve the QZD. 

In addition, we have observed that the frequency of the oscillating leakage decreases when \(\delta\) increases. We now give an explanation. \(U^{(1)}(\tau)\ket{1}\) in \Eq{eq:ut1_after_plugin_matrix_ele} is a superposition of oscillations. The terms with the largest oscillation amplitude in \Eq{eq:ut1_after_plugin_matrix_ele} are:
\al{
\ket{\phi_{\frac{N-2}{2}}^{(0)}}&\smlb{e^{-i{\eta}_{\frac{N-2}{2}}\tau}-e^{-i{\eta}_{0\alpha}\tau}}\dfrac{G}{\sqrt{2}}\nonumber\\&-\ket{\phi_{\frac{N}{2}}^{(0)}}\smlb{e^{-i{\eta}_{\frac{N}{2}}\tau}-e^{-i{\eta}_{0\beta}\tau}}\dfrac{G}{\sqrt{2}},
\label{eq:largestterm}
}
assuming \((N/2-1)\) is odd. Due to the symmetry of \(H_\text{tot}\), \({\eta}_{\frac{N-2}{2}}=-{\eta}_{\frac{N}{2}}\) and \({\eta}_{0\alpha}=-{\eta}_{0\beta}\). Therefore, the total time dependence of the norm of \Eq{eq:largestterm} is given by
\al{
&\sqrt{\abs{e^{-i{\eta}_{\frac{N-2}{2}}\tau}-e^{-i{\eta}_{0\alpha}\tau}}^2+\abs{e^{+i{\eta}_{\frac{N-2}{2}}\tau}-e^{+i{\eta}_{0\alpha}\tau}}^2}\nonumber\\
&=2\sqrt{1-\cos{[\lambda^{-1}(\eta_{\frac{N-2}{2}}-{\eta}_{0\alpha})t]}}.
}
According to \Eq{eq:etan}, increasing N or decreasing \(\lambda^{-1}\) will decrease the difference between \(\lambda^{-1}\eta^{(0)}_{\frac{N-2}{2}}\) and \(\lambda^{-1}\eta^{(0)}_{0\alpha}=0\). This also implies a decrease in the frequency \(\lambda^{-1}(\eta_{\frac{N-2}{2}}-{\eta}_{0\alpha})\), which then infers a decrease in the frequency of \(\abs{U^{(1)}(t)\ket{1}}^2\), and finally a decrease in the frequency of leakage. This explains why increasing N or decreasing \(\lambda^{-1}\) increases leakage's amplitude and decreases its frequency simultaneously.
\newline
\newline
\newline
\section{Derivation of \Eq{eq:evenN}}
\label{sec:evenH2}

We first write \(H_\text{tot}\) \Eq{eq:hChain} in a matrix form:
\begin{widetext}
\al{
H_\text{tot}=\lambda^{-1}H_w+H =\lambda^{-1}\begin{bmatrix} 
     0 & 0 &  &  &  &  & &  & \\
     0 & 0 & k &  &  & & \text{\huge0} & &  \\
      & k & 0 & k &  & & & &  \\
     &  & k & 0 & k & &  &  & \\
     &  &  & \ddots & \ddots & \ddots&  & & \\
     &  &  &  &  &  &  & k & \\
     &  & \text{\huge0}  &  &  & &  k & 0 & 0\\
     &  &  &  &  &  &   & 0 & 0
    \end{bmatrix}_{N\times N}+
\begin{bmatrix}
  \begin{pmatrix}
  0 & k \\
  k & 0
  \end{pmatrix} && & &\\
 & & & &\\
& & \text{\huge0} & &\\
 & & & & \\
 & & & & \begin{pmatrix}
  0 & k \\
  k & 0
  \end{pmatrix}
\end{bmatrix}_{N\times N}.
\label{eq:hmatrix}
}
\end{widetext}
We define a unitary N-by-N matrix
\al{
R=\smlb{\ket{\phi_{0\alpha}^{(0)}},\ket{\phi_{1}^{(0)}},..., \ket{\phi_{N-2}^{(0)}},\ket{\phi_{0\beta}^{(0)}}},
}
which diagonalizes \(H_w\), 
\al{
R^\dagger H_w R=\Lambda\equiv\text{diag}\{0,\eta^{(0)}_1,\eta^{(0)}_{2},..., \eta^{(0)}_{N-2},0\}.
\label{eq:diag_Hm}
}
For future use, because the boundary elements (first and last rows/columns) in \(H_w\) are all zero, we define
\al{
{R'}^\dagger H'_w {R'}=\Lambda'\equiv\text{diag}\{\eta^{(0)}_1,\eta^{(0)}_{2},..., \eta^{(0)}_{N-2}\}.
\label{eq:diag_H'm}
}
where (N-2)-by-(N-2) matrices \(H'_w\), \(R'\) and \(\Lambda'\) are obtained from \(H_w\), \(R\) and \(\Lambda\) by stripping off their boundary elements respectively. 

Given that N is even, there is no third zero in \(\Lambda\) \Eq{eq:diag_Hm} according to \Eq{eq:etan}. This entails \(P^{(0)}_0=\ket{\phi_{0\alpha}^{(0)}}\bra{\phi_{0\alpha}^{(0)}}+\ket{\phi_{0\beta}^{(0)}}\bra{\phi_{0\beta}^{(0)}}=\ket{1}\bra{1}+\ket{N}\bra{N}\). Inserting it into \(H_{\text{QZD}}^{(1)}\), we obtain
\al{
\begin{aligned}
H_{\text{QZD}}^{(1)}&=P^{(0)}_{0}H{\widetilde{Q}^{(0)}_0}HP^{(0)}_{0}\\
&=k^2\brak{2}{{\widetilde{Q}^{(0)}_0}}{N-1}\ket{1}\bra{N}+h.c.\\
&\quad+{k}^2\brak{2}{{\widetilde{Q}^{(0)}_0}}{2}\ket{1}\bra{1}\\
&\quad+{k}^2\brak{N-1}{{\widetilde{Q}^{(0)}_0}}{N-1}\ket{N}\bra{N}, 
\end{aligned}
\label{eq:coup_mat}
}
where
\[
{\widetilde{Q}^{(0)}_0}=\sum_{n=1}^{N-2}\dfrac{\ket{\phi_{n}^{(0)}}\bra{\phi_{n}^{(0)}}}{-\eta^{(0)}_n}.
\]
Therefore, the matrix element in \Eq{eq:coup_mat} can be derived as
\al{
\brak{i}{{\widetilde{Q}^{(0)}_0}}{j}=-\brak{i}{R'{\Lambda'}^{-1}{R'^\dagger}}{j}=-\bra{i}{(H'_w)}^{-1}\ket{j},
}
where we have used
\al{
&{\Lambda'}^{-1}=\midb{{R'}^\dagger H'_w {R'}}^{-1}\nonumber
\\\Longrightarrow &{R'}\Lambda^{-1}{R'}^\dagger=\cancel{{R'}{R'}^\dagger}{(H'_w)}^{-1}\cancel{{R'}{R'}^\dagger}.\nonumber
}
A mathematical work \cite{usmani1994inversion} gives the analytical expression of the elements in the inverse of a general tridiagonal matrix. Using the expression, we obtain \(\bra{2}{\widetilde{Q}^{(0)}_0}\ket{2}=\bra{2}{(H'_w)}^{-1}\ket{2}=0\), \(\bra{N-1}{\widetilde{Q}^{(0)}_0}\ket{N-1}=\bra{N-1}{(H'_w)}^{-1}\ket{N-1}=0\) and \(\bra{2}{\widetilde{Q}^{(0)}_0}\ket{N-1}=-\bra{2}{(H'_w)}^{-1}\ket{N-1}=(-1)^{\frac{N}{2}-1}/{k}\). Inserting them into \Eq{eq:coup_mat}, we derive \Eq{eq:evenN} in the main text,
\[
\lambda H_{\text{QZD}}^{(1)}(\text{even}~N)=(-1)^{\frac{N}{2}-1}\lambda k\smlb{\ket{1}\bra{N}+h.c.}.
\]

To consider a realistic tight-binding chain, we allow for a small random fluctuation in the coupling intensity, forming a new matrix
\al{
H_{w,\text{fluc}}=\sum_{i=2}^{N-2} k_i\smlb{\ket{i}\bra{i+1}+\textit{h.c.}},
}
where \(k_i\approx k\). The new matrix also yields \(\bra{2}{\widetilde{Q}^{(0)}_0}\ket{2}=\bra{N-1}{\widetilde{Q}^{(0)}_0}\ket{N-1}=0\). On top of that, 
\al{
\bra{2}{\widetilde{Q}^{(0)}_0}\ket{N-1}=(-1)^{\frac{N}{2}-1}\dfrac{\prod_{\text{odd}~i}k_{i}}{\prod_{\text{even}~i}{k_{i}}}&\approx(-1)^{\frac{N}{2}-1}\dfrac{k^{\frac{N}{2}-2}}{k^{\frac{N}{2}-1}}\nonumber\\&=(-1)^{\frac{N}{2}-1}/k,
}
where fluctuations in \(k_i\) averages out through products. Therefore, the results have little difference from before. This means that the form of \(H_{\text{QZD}}^{(1)}(\text{even}~N)\) does not rely on the identical coupling intensity, and thus shows the viability of experimental realization.

\section{Why a modified odd chain retrieves FC-QZD}
\label{sec:odd_detune}
Unlike even chains, odd chains has \(\eta_{(N-1)/2}^{(0)}=0\) except for \(\eta_{0}^{(0)}=0\) according to \Eq{eq:etan}, whose corresponding eigenstates is \(\ket{\phi_{0\text{mid}}}\) in \Eq{eq:phi_mid}. This entails \(P^{(0)}_0=\ket{1}\bra{1}+\ket{N}\bra{N}+\ket{\phi_{0\text{mid}}}\bra{\phi_{0\text{mid}}}\), which has been shown to ruin FC-QZD in \Sec{sec:chain}. Therefore, reviving FC-QZD requires excluding any zero eigenvalue except for \(\eta_{0}^{(0)}=0\).

To this end, we choose to construct \Eq{eq:hmd}
\(
H_\text{mod}=H_{\text{tot}}+\Delta\omega\ket{2}\bra{2}=H+\lambda^{-1}(H_w+\lambda\Delta\omega\ket{2}\bra{2})\). This modification is meant to ensure \(\text{det}(H'_w+\lambda\Delta\omega\ket{2}\bra{2})=(-1)^{\smlb{N-3}/2}k^{N-3}\lambda\Delta\omega\neq0\). The equation is derived according to \cite{el2004inverse}, which provides a general formula for the determinant of tridiagonal matrices. The nonzero determinant is equal to the product of all eigenvalues after modification except for \(\eta_{0}^{(0)}=0\). (Before modification, \(\text{det}(H'_w)=\prod_{n\neq0} \eta_n^{(0)}=0\).) Therefore, we conclude that there is no other zero eigenvalue except for \(\eta_{0}^{(0)}=0\) after modification. Eventually, we retrieve \(P^{(0)}_0=\ket{1}\bra{1}+\ket{N}\bra{N}\) and FC-QZD. In fact, the formula in \cite{el2004inverse} suggests that, a modification with \(\ket{2}\) replaced by any other even site is also effective. 

Then, by repeating the procedure in \App{sec:evenH2} with \(H_\text{tot}\) replaced by \(H_\text{mod}\), we obtain \Eq{eq:odd_mod} in the main text.

\bibliography{main}
\end{document}